\documentclass{PoS}
\usepackage{epsfig}
\usepackage{latexsym}
\usepackage{amsmath}
\usepackage{amsfonts}
\newcommand{\rmii}[1]{{\mbox{\tiny\rm{#1}}}}
\newcommand{\Tr}{\mathop{\textrm{Tr}}}
\newcommand{\Nc}{N_{\rm c}}
\def\lsi{\raise0.3ex\hbox{$<$\kern-0.75em\raise-1.1ex\hbox{$\sim$}}}
\def\gsi{\raise0.3ex\hbox{$>$\kern-0.75em\raise-1.1ex\hbox{$\sim$}}}
\newcommand{\lsim}{\mathop{\lsi}}

\title{Towards 4-loop NSPT result for a 3-dimensional condensate-contribution to hot QCD pressure}

\ShortTitle{Towards 4-loop NSPT result for a 3d condensate-contribution to hot QCD pressure}

\author{\speaker{C. Torrero}, M. Laine, Y. Schr\"oder \\

        Faculty of Physics, University of Bielefeld, 
        33501 Bielefeld, Germany
\\

        E-mail: \email{torrero@physik.uni-bielefeld.de,\\ laine@physik.uni-bielefeld.de,\\ yorks@physik.uni-bielefeld.de}}

\author{F. Di Renzo\\

        Università di Parma \& INFN, Parco Area delle Scienze 7A, 
        43100 Parma, Italy
\\

        E-mail: \email{direnzo@fis.unipr.it}}

\author{V. Miccio\\

        CERN, IT/PSS,
        1211 Gen\`eve 23, Switzerland 
\\

        E-mail: \email{vincenzo.miccio@cern.ch}}

\abstract{Thanks to dimensional reduction, the contributions to the hot QCD
pressure coming from so-called soft modes can be studied via an effective
three-dimensional theory named Electrostatic QCD (spatial Yang-Mills
fields plus an adjoint Higgs scalar). The poor convergence of the perturbative
series within EQCD suggests to perform lattice measurements of some of 
the associated gluon condensates. These turn out, however, to be plagued 
by large discretization artifacts. We discuss how Numerical Stochastic 
Perturbation Theory can be exploited to determine the full lattice spacing 
dependence of one of these condensates up to 4-loop order, and sharpen our tools 
on a concrete 2-loop example.}

\FullConference{The XXV International Symposium on Lattice Field Theory\\
                 July 30-4 August 2007\\
                 Regensburg, Germany}

\begin{document}

\section{Motivation}

The pressure $p(T)$ of hot QCD (with a temperature $T \gg 150$~MeV) 
is of crucial importance in different contexts. For instance, in \emph{cosmology}, 
the cooling rate of the Universe is given by
%
\begin{equation}
 \frac{1}{T}\frac{{\rm d}T}{{\rm d}t} = - \frac{\sqrt{24\pi}}{m_\rmii{Pl}}
 \frac{\sqrt{e(T)} s(T)}{c(T)} \;,
\end{equation}
where $T$ is the temperature, $t$ the age of the Universe, 
$m_\rmii{Pl}$ the Planck mass, and
%
\begin{equation}
 s(T)=p'(T)~, ~~~~ e(T)=Ts(T)-p(T)~, ~~~~c=e'(T)~.
\end{equation}
On the other hand, 
in \emph{heavy ion collisions}, the system evolves according to
%
\begin{equation}
T^{\mu\nu} \approx [p(T)+e(T)]u^{\mu}u^{\nu}-p(T)\eta^{\mu\nu}~, ~~~~\partial_{\mu}T^{\mu\nu}=0~,
\end{equation}
where $T^{\mu\nu}$ is the energy-momentum tensor,  $\eta^{\mu\nu}$ the Minkowski metric,
and $u^{\mu}$ the flow velocity.
Finally, from \emph{a theoretical point of view}, 
the pressure is proportional to the number of effective degrees of freedom, 
and is therefore a good observable to characterize hot QCD matter.


\section{Theoretical setup --- part I (continuum)}

In general, the determination of the QCD pressure is a non-trivial task.
In spite of our restriction to the deconfined phase, $T \gg 150$~MeV, 
where perturbation theory should in principle
be applicable, it is of limited use in practice, because of the very slow 
convergence of the perturbative series~\cite{Arn}. At the same time, first principles
lattice simulations are also difficult at temperatures above about $T \sim 1$~GeV~\cite{Boy}, 
because the system then develops a scale hierarchy, $g^2T/\pi \lsim gT \lsim \pi T$, 
where $g$ is the QCD gauge coupling (for an idea to possibly overcome this
limitation, see ref.~\cite{Fod}).

A suitable strategy to tackle the computation of $p(T)$ 
(and of many other observables) in this situation
is given by \emph{Dimensional Reduction}~\cite{Gin,App,Kaj}:
it consists of integrating out ``hard modes'', with momenta $k\sim \pi T$,   
from the four-dimensional (4d) QCD,  
to arrive at an effective description 
in terms of so-called \emph{Electrostatic QCD (EQCD)}~\cite{Bra},
i.e.\ a three-dimensional (3d) Yang-Mills theory plus an adjoint Higgs field $A_0^a(x)$.
The action of EQCD is given by 
%
\begin{eqnarray}
 S_\rmii{EQCD}&=&\int \! {\rm d}^3x \,
 \biggl\{\frac{1}{2}\Tr[F_{ij}^{2}(x)]+\Tr[D_i,A_0(x)]^{2}+
 m_\rmii{E}^{2}\Tr [A_0^{2}(x)]+
 \lambda_\rmii{E}\big(\Tr[A_0^{2}(x)]~\!\big)^{2}\biggr\}~,\;
\end{eqnarray}
where ${F}_{ij}^{~\!\!a}$ is the 3d field strength tensor; 
$D_i$ the covariant derivative; $A_0=\sum_{B=1}^{8}A_0^{B}~\!T^{B}$ 
with $\Tr~\![T^{A}T^{B}]=\frac{1}{2}\!\delta^{AB}$;
and we implicitly assume the use of dimensional regularization. 
The temperature $T$ now enters 
only via the parameters $m_\rmii{E}$, $\lambda_\rmii{E}$ 
and $g_\rmii{E}$, where $g_\rmii{E}$ is the EQCD gauge coupling.

It turns out that perturbation theory within EQCD might converge very slowly~\cite{Bra} (see, however, ref.~\cite{conv}).
One can then switch to a numerical measurement of the partial derivatives of the pressure with respect to 
the parameters of the EQCD action, the so-called \emph{condensates}: after subtracting the proper counterterms, 
the results are extrapolated to the continuum and then numerically integrated to finally get $p(T)$~\cite{Kaj2}. 
It turns out that carrying out the continuum extrapolation is difficult because of large $O(a)$ discretization effects: 
thanks to super-renormalizability, \emph{these effects are however purely perturbative in nature}, 
and our aim is to determine them up to 4-loop level.


\section{Theoretical setup --- part II (lattice)}

The EQCD action on the lattice may be written as
\vspace*{-0.5cm}

\begin{eqnarray}
S_\rmii{latt}&=&\beta\sum_\rmii{$x,~\!i<j$}\Bigl(1-\frac{1}{3}\mathop{\rm Re}\Tr\big[P_{ij}(x)\big]\Bigr) - \nonumber\\
&-&2\sum_\rmii{$x,~\!i<j$}\Tr\big[\phi(x)~\!U_i(x)~\!\phi(x+i)~\!U_i^{\dag}(x)\big] + \nonumber\\
&+&\sum_\rmii{$x$}\bigg\{\alpha(\beta,\lambda, am_\rmii{E})\Tr\big[\phi^2(x)\big]+\lambda\Big(\!\Tr\big[\phi^2(x)\big]\Big)^2 \bigg\}~,
\end{eqnarray}
where $\beta=6/a g_\rmii{E}^2$, $P_{ij}$ is the plaquette,
$U_i$ is the link variable, $\phi=\sqrt{a} A_0$,
$\lambda = a \lambda_\rmii{E}$, and~\cite{Lai}
\vspace*{-0.25cm}

\begin{eqnarray}
\alpha(\beta,\lambda,am_\rmii{E})&=&
6\bigg\{1+\frac{(am_\rmii{E})^2}{6} -\bigg(6+\frac{5}{3}\lambda\beta\bigg)
\frac{3.175911525625}{4\pi\beta}-\nonumber\\
&-&\frac{3}{8\pi^{2}\beta^{2}}~\!
\bigg[\bigg(10\lambda\beta-\frac{5}{9}\lambda^{2}\beta^{2}\bigg)\bigg(\ln\beta+0.08849\bigg)+
\frac{34.768}{6}\lambda\beta+36.130\bigg]\bigg\}~. \nonumber \\
\end{eqnarray}
The quantity under inspection is the derivative of $p(T)$ 
with respect to $(am_\rmii{E})^2$;
this yields $\langle~\!\! \Tr~\!\![~\!\phi^2]~\!\rangle$, which can be expanded as
\vspace*{-0.5cm}

\begin{eqnarray}
 \langle~\!\! \Tr~\!\![~\!\phi^2]~\!\rangle&=&
 d_{00}+d_{10}~\!\frac{1}{\beta}+d_{11}\lambda+
 d_{20}~\!\frac{1}{\beta^{2}}+d_{21}~\!\frac{\lambda}{\beta}+d_{22}\lambda^2+\nonumber\\
 &+&d_{30}~\!\frac{1}{\beta{^{3}}}+
    d_{31}~\!\frac{\lambda}{\beta{^{2}}}+
    d_{32}~\!\frac{\lambda^2}{\beta}+
    d_{33}\lambda^3+{O}\bigg(~\!\frac{\lambda^n}{\beta{^{4-n}}}\bigg)~.
\vspace{0.5cm}
\end{eqnarray}
The coefficients $d_{00}$, $d_{10}$, $d_{11}$, $d_{21}$ and $d_{22}$ 
are  known analytically, for instance ($N\equiv$  lattice extent)
\begin{eqnarray}
 d_{11} & = &  40 \biggl[ \frac{3.175911525625}{4\pi}
 - 
 \frac{1}{N^3} \sum_{n_1 = 0}^{N-1}\,\sum_{n_2 = 0}^{N-1}\,\sum_{n_3 = 0}^{N-1} 
 \frac{1}{4 \sum_{i=1}^3 \sin^2(\frac{\pi n_i}{N}) + (a m_\rmii{E})^2 }
 \biggr]
 \times \nonumber \\  & \times & 
 \frac{1}{N^3} \sum_{k_1 = 0}^{N-1}\,\sum_{k_2 = 0}^{N-1}\,\sum_{k_3 = 0}^{N-1} 
 \biggl[ \frac{1}{4 \sum_{i=1}^3 \sin^2(\frac{\pi k_i}{N}) + (a m_\rmii{E})^2} \biggr]^2
 \;, \label{d11}
\end{eqnarray} 
but the others have to be determined: those with the biggest impact 
are expected to be the 3-loop and 4-loop coefficients 
independent of $\lambda$, i.e.\ ${d_{20}}$ and ${d_{30}}$, 
respectively.


\section{Numerical setup}

The perturbative study is concretely carried out by means of \emph{Numerical Stochastic Perturbation Theory (NSPT)}~\cite{FDR}.
(Incidentally, it would also be interesting to pursue 
the same computation with standard techniques~\cite{HP}.)
Its origins lie in {Stochastic Quantization}~\cite{Par}, based on introducing an extra coordinate $t$
and {an evolution equation of the Langevin type}, namely
%
\begin{equation}
 \partial_t\phi(x,t)= -\partial_{\phi} S[\phi]+\eta(x,t)~,
\end{equation}
where $\eta(x,t)$ is a Gaussian noise.
The usual Feynman-Gibbs path integral is recovered by averaging over the stochastic time,
%
\begin{equation}
 Z^{-1}\!\!\int [D\phi]
 O[\phi(x)]e^{-S[\phi(x)]}=\lim_{t\rightarrow\infty}
 \frac{1}{t} \int_0^t \! {\rm d} t' \,
 \big\langle O[\phi_{\eta}(x,t')]\big\rangle_{\eta}\;.
\end{equation}
NSPT can now be introduced by expanding the variables as
\begin{equation}
 \phi(x,t)\longrightarrow\sum_{k}g_0^k\phi^{(k)}(x,t)\;,
\end{equation}
where $g_0$ is some small coupling. 
This results in a hierarchical system of differential equations that can be numerically integrated by discretizing
the stochastic time, as $t = n\tau$, where $\tau$ is a time step. 

A similar construction holds also for the gauge degrees of freedom $U_i(x)$, for which the Langevin equation reads
%
\begin{equation}
 \partial_{t}U_{\eta} = -i\Bigl( \nabla S[U_{\eta}]+\eta \Bigr)
 U_{\eta} \;,
\end{equation}
in order to assure a correct evolution within the group. The perturbative expansion is then
a double expansion in $\beta$ and $\lambda$, to obtain the previously-written
series of $\langle~\!\! \Tr~\!\![~\!\phi^{2}]~\!\rangle$.

In practice, every variable evolves according 
to its Langevin dynamics for different values of $\tau$;
a measurement of $\Tr~\!\![~\!\phi^{2}]$ is performed at every time step
(once thermalization has been reached); and finally one extrapolates to $\tau = 0$ 
(this last step is necessary since the correct probabilistic weight at equilibrium is recovered 
only in the limit $\tau \rightarrow 0$).
This procedure is then repeated after changing the parameters of the action.

%
\medskip
\begin{figure}
\centerline{\includegraphics[height=0.48\textwidth,width=0.48\textwidth]{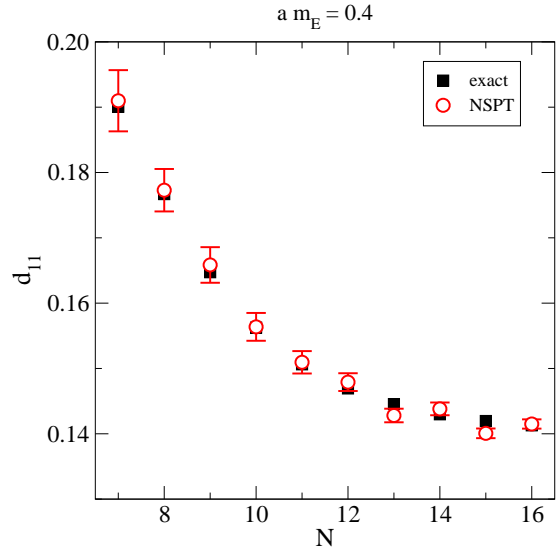}}
\caption{\em Analytical and numerical results for the coefficient $d_{11}$ vs.\ $N$ at $a m_\rmii{E} = 0.4$. 
The results agree with each other within statistical errors.}
\label{Fig.1}
\end{figure}
%


\section{Preliminary results}

Our approach involves three different extrapolations / interpolations in total: 
first, the above-mentioned extrapolation to $\tau\to 0$; second, an extrapolation to infinite volume ($N\to\infty$); 
third, an interpolation between the different $a m_\rmii{E}$ simulated, 
in order to obtain the desired 4-loop predictions at any finite $a m_\rmii{E}$.

As a check of the first of these extrapolations, we compare the numerical outputs 
for $d_{11}$ with its known values 
(eq.~(\ref{d11})) at fixed $a m_\rmii{E}$ and lattice extent $N$: 
this is done in {\bf Fig.~1}.

The next task is to extrapolate to infinite volume. As Fig.~1 shows, finite-volume effects become exponentially
small at large volumes. However, the volume required for this grows as the mass $a m_\rmii{E}$ decreases. This is
illustrated in {\bf Fig.~2} for two masses, one larger than in Fig.~1, and one smaller. For the large mass, a plateau can
be reached allowing for a reliable infinite-volume extrapolation, while for the smaller mass, the largest volumes
we can afford are not yet large enough to reach a plateau.

In order to deal with this situation, we adopt the following approach.
Let us consider a mass like $a m_\rmii{E} = 0.8$: numerical evidence shows that
this one has a reasonable plateau at affordable $N$ for all the coefficients $d_{ij}$,
but still the behaviour of the data is not too flat (i.e., some volume dependence is detectable).
One can then extract an infinite-volume value $d_{ij}(\infty)$ by fitting a constant to
data in the range of the plateau, and
subtract it from the data in order to obtain the quantities ($L \equiv N a$)

%
\medskip
\begin{figure}
\centerline{\includegraphics[height=0.48\textwidth, width=0.48\textwidth]{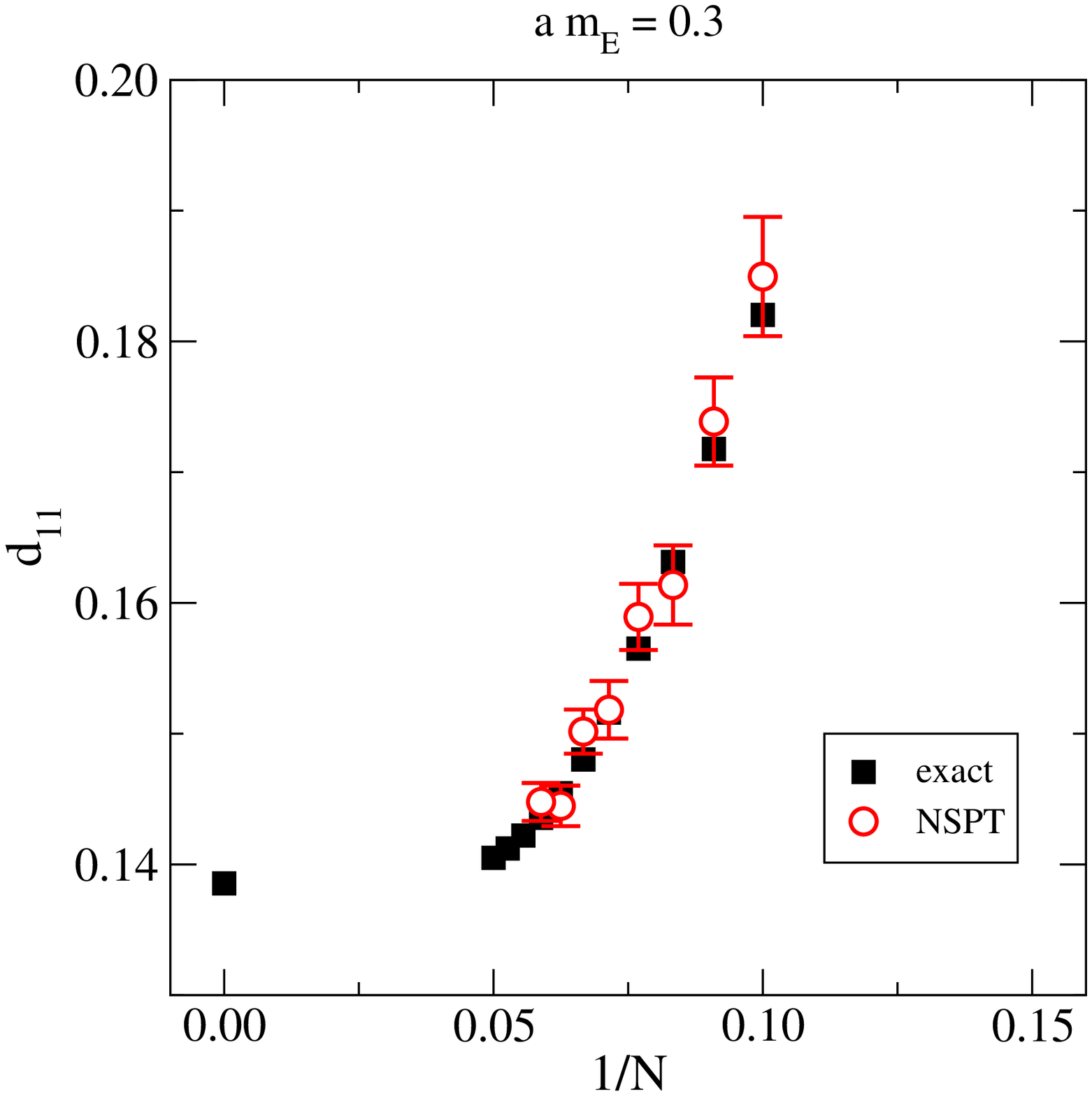}%
            ~~~ \includegraphics[height=0.48\textwidth, width=0.48\textwidth]{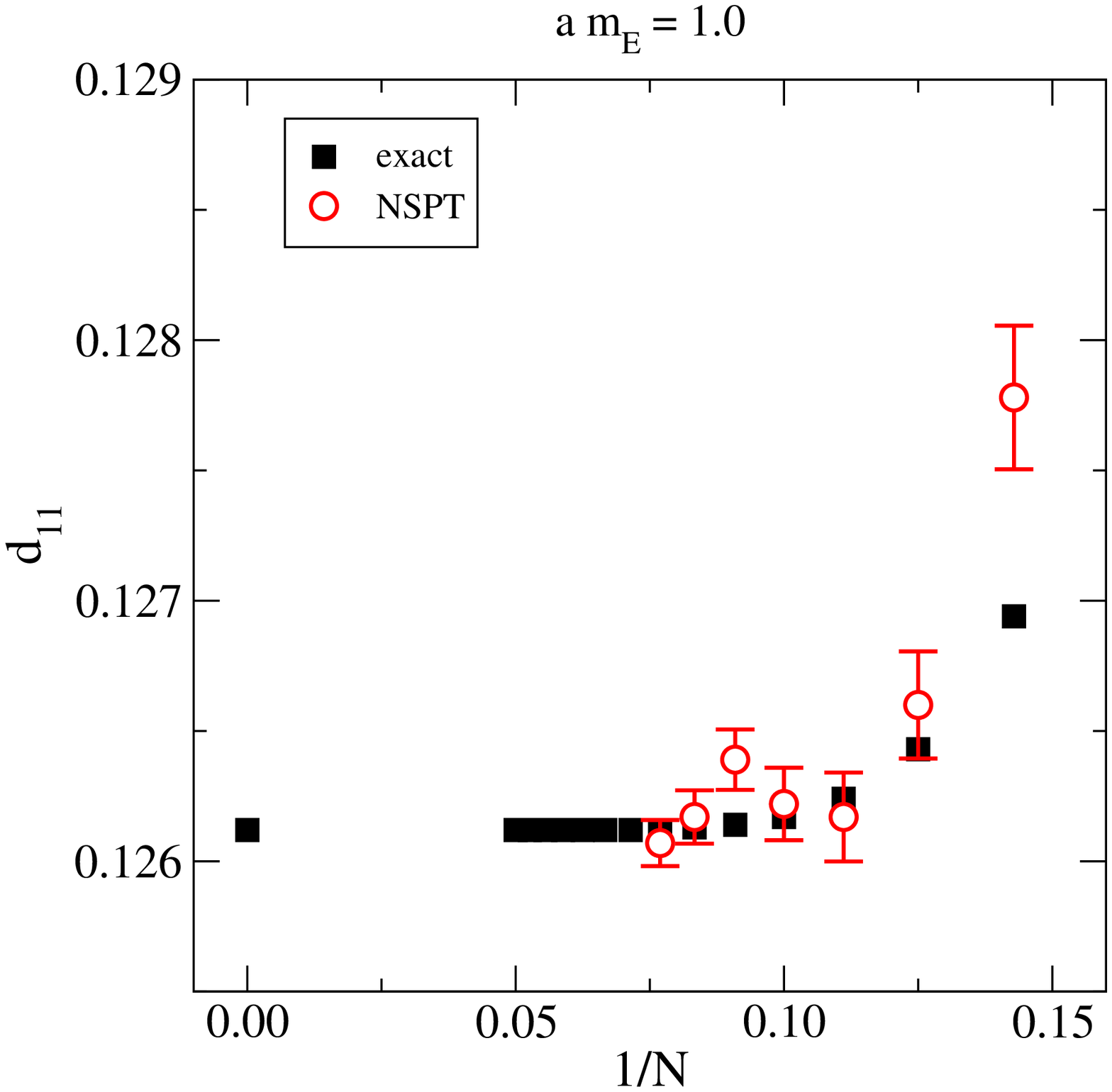}}
\caption {\em Results for the coefficient $d_{11}$ vs.\ $1/N$ at $a m_\rmii{E} = 0.3$ and $1.0$.
Note the very different resolutions of the vertical axes.}
\label{Fig.2}
\end{figure}
%

\vspace{-0.3cm}
\begin{equation}
 g_{ij}(m_\rmii{E}L) \equiv d_{ij}(m_\rmii{E}L) - d_{ij}(\infty)~.
\vspace{0.2cm}
\end{equation}
Subsequently, one can try to obtain a reasonable interpolating 
fit $f_{ij}(m_\rmii{E}L)$ for $g_{ij}(m_\rmii{E}L)$, allowing to go also
to other values of $m_\rmii{E} L$ than those simulated at $a m_\rmii{E} = 0.8$.
After this, one can go back to the other masses $m'_\rmii{E}$, 
and take a finite-size scaling ansatz of the form
%
\begin{equation}
d_{ij}(m'_\rmii{E}L) = d_{ij}(\infty) + A_{ij}(a m'_\rmii{E}) \times f_{ij}(m'_\rmii{E}L)~,
\end{equation}
where $d_{ij}(m'_\rmii{E}L)$ are the direct measurements at the mass $m'_\rmii{E}$,
and $d_{ij}(\infty)$ and $A_{ij}(a m'_\rmii{E})$ are
volume-independent fit coefficients. Test results for $d_{11}(\infty)$
obtained this way are shown in {\bf Fig.~3}.

%
\medskip
\begin{figure}
\centerline{\includegraphics[height=0.48\textwidth, width=0.48\textwidth]{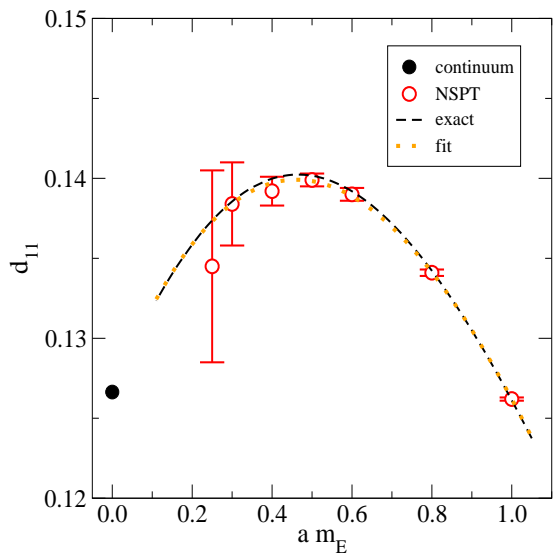}}
\caption{\em Comparison between exact values (dashed curve) 
and numerical infinite-volume extrapolations (open symbols) 
as a function of $a m_\rmii{E}$.
The continuum value at $am_\rmii{E} = 0$ has been extracted from ref.~\cite{Kaj3}.
The dotted line shows a fit through the open symbols, constrained to go through
the continuum point (which is known for all $d_{ij}$).  
The agreement between the exact and fitted curves is satisfactory. 
}
\label{Fig.3}
\end{figure}
%


\section{Conclusions}

Preliminary studies of the 2-loop coefficient $d_{11}$, where analytical values
are also available, show that our general approach works. The next goal is to finalise the analysis
for the most important 3-loop coefficient $d_{20}$ and 4-loop coefficient
$d_{30}$~\cite{Prog}. The quality of our data is good, so we expect to be able to extract these with small
errors in the same mass range as in Fig.~3. This should allow to re-analyse the Monte Carlo
data of ref.~\cite{Kaj2} with significantly reduced systematic errors from
the continuum extrapolation. Combining with refs.~\cite{gsixg,Hie,FDR2},
all ``soft'' contributions to the hot QCD pressure would 
then be under reasonable control. At the same time, 
the determination of the 4-loop ``hard''  contributions remains an open challenge; 
toy model computations in scalar field theory
have suggested, however, that it \emph{can} be tackled with some effort~\cite{phi4}. 


\section*{Acknowledgements}

We warmly thank {\em ECT*, Trento,} for providing computing time 
on the {\em BEN} system. This work was partly supported by the DFG 
project \emph{Precision physics with hot QCD}.


\end{document}